# Application of Modified Multi Model Predictive Control Algorithm to Fluid Catalytic Cracking Unit

Nafay H. Rehman[1], Neelam Verma[2]

*Student[1], Asst. Professor[2]*

Department of Electrical and Electronics Engineering

Amity University, Noida, India

*Abstract*—This paper presents a modified multi model predictive control algorithm for the control of riser outlet temperature and regenerator temperature for the fluid catalytic cracking unit (FCCU). The models of the fluid catalytic cracking unit are estimated using subspace identification (N4SID) algorithm. The PRBS signal is applied as an input signal to estimate the FCCU models. Since the estimated model does not give 100% fit; especially for nonlinear systems having more than one operating conditions, multi-model approach is proposed. In multi model, more than one model of FCCU used in MPC design. The main advantages of proposed method are that it can handle hard input and output constraints and it can be used for multi input multi output processes (MIMO) without increasing the complexity in control design. MATLAB/Simulink is used to estimate the models of FCCU and simulate the results for the controller. The simulation results show that the proposed algorithm provides better result for both reference tracking and disturbance rejection.

## I. INTRODUCTION

System identification is a process to build a mathematical model of dynamical systems from observed input and output data. These methods are widely used in industry for over a decade. It is being used in process control, aerospace, automotive, disk drives and embedded systems to create system models for any systems. The advantage of system identification is that it is quick and applicable to almost all systems [Billings, 2013].

On another hand model predictive control (MPC) is an advanced optimized control method that has been widely used in process industries over the last two decades. It is a form of control in which the model of the process being controlled is required. The controller uses the model of the process and the output measurements to calculate the current control actions and predict the future behavior of the processes. The control action is calculated by minimizing the cost function at each sampling instants.

This paper proposes the modified model predictive control algorithm based on system identification and model predictive control approach.

## II. SYSTEM IDENTIFICATION

The procedure to identify the dynamical model for any systems is as follows. First we have to select an input signal and apply it to the plant to collect an output data. Input signal can be impulse signal, step signal, pseudo random binary noise (PRBS), generalized binary noise (GBN), multiple sinusoids, etc. After getting the input and output data, the model structure of the plant is specified. There are three common types of models in system identification: white-box identification model, black-box identification model and grey-box identification model. White-box identification model structure is based on the first principles such as Newton's law. In white-box identification; the model structure is completely known and the model parameters are estimated from the measured data. In grey-box identification the model structure is partially known from the first principles and the rest is developed from the measured data. In black-box identification, the model structure and its parameters are completely unknown and they are estimated using observed input and output data. After deciding the model structure, the system identification algorithms is used to estimate the mathematical models of dynamical systems. After the mathematical models of dynamical systems are developed, the result is validated. If the estimated model is not good enough then other estimation methods can be tried.

Input signals play very important role in system identification because it is the only way to check the behavior of the process and collect the output data. In general we cannot introduce any random input to the process that is being estimated. We have to select an input signal that will carry enough knowledge about the system and affect smoothly all the operating frequencies. This can be achieved by using many different excitation signals, such as impulse signals, step signals, pseudo random binary sequences (PRBS), generalized binary noises (GBN) and multiple sinusoids. The choice among these input signals depends on the type of identification technique used and the priori knowledge of the system under the test.

In the proposed method, PRBS signal is used as an input signal. PRBS is a two stage deterministic signal with a periodic sequence of length 'N' that switches between two levels, e.g. $+a$ and $-a$. To generate these sequences, there are two possibilities: first possibility is to use a quadratic residue code method suggested by Godfrey [Godfrey, 1993] and the second possibility is to use feedback shift register [Godfrey, 1993] and [Eykhoff, 1974].

PRBS signals have been used for nonparametric model identification, such as: frequency response estimation and correlation analysis. In process control white noise signal is harmful to the actuator because it over emphasize the high frequency band at the cost of the low and the middle frequency band. In process control low pass character signals are preferred. Such signals can be obtained by increasing the clock period of the signal or by filtering the PRBS signals. The PRBS signal is preferred for system identification, because it excites all frequencies equally and imitate white noise in deterministic signal [Nelles, 2011].

### III. N4SID Subspace Identification

The state space representation of dynamical system given as
$$x(k+1) = Ax(k) + Bu(k) + Ke(k) \quad (1)$$
$$y(k) = Cx(k) + Du(k) + e(k) \quad (2)$$
where $x(k)$ is the state vector, $y(k)$ is the system output, $u(k)$ is the system input, $e(k)$ is the stochastic error, A, B, C, D and K are the system matrices.

Subspace identification is a method used to estimate the state space matrices A, B, C and D from an input and output data. It was proposed by van Overschee and de Moor. Further development to this method was done by Larimore in 1990 in which he proposed canonical variate analysis (CVA) [Larimore, 1990].

In 1994, van Overschee and de Moor proposed new numerical algorithm N4SID which identifies the mixed deterministic-stochastic systems.

In N4SID, the observability and controllability of the system is not needed to be known in advance since the state space matrices are not calculated in their canonical form but it is calculated as the full state space matrices so there is no problem of identifiability [Overschee and Moor, 1994].

Let consider a state space model of combined deterministic-stochastic system given as
$$x_{k+1} = Ax_k + Bu_k + w_k \quad (3)$$
$$y_k = Cx_k + Du_k + v_k \quad (4)$$
where A, B, C and D are the state space matrices, $w_k$ is the state noise with covariance matrices $E[v_k v_k^T] = Q$ and $v_k$ is the output measurement noise with covariance matrices $E[w_k w_k^T] = R$ and $E[v_k w_k^T] = S$.

If the system is observable then Kalman filter can be designed for the system to estimate the state variables according to
$$\hat{x}_{k+1} = A\hat{x}_k + Bu_k + K(y_k - C\hat{x}_k - Du_k) \quad (5)$$
where K is the steady state Kalman gain that can be solved using Ricatti equation.
$$P = APA^T - K(C^T PC + R)K^T + Q \quad (6)$$
Denoting
$$e_k = y_k - C\hat{x}_k - Du_k \quad (7)$$
Substituting (7) in (5) gives (1) and (2).

The system described by (3) and (4) can be represented in the predictor form as
$$x_{k+1} = A_K x_k + B_K z_k \quad (8)$$
$$y_k = Cx_k + Du_k + e_k \quad (9)$$
where
$$z_k = [u_k^T, y_k^T], A_K = A - KC, B_K = [B - KD, K] \quad (10)$$
From (8) and (9), an extended state space can be formulated as
$$y_f(k) = \Gamma_f x_k + \overline{G}_f z_{f-1}(k) + D_f u_f(k) + e_f(k) \quad (11)$$
where f is the future horizon and
$$\Gamma_f = \begin{bmatrix} C \\ CA_K \\ \vdots \\ CA_K^{f-1} \end{bmatrix}, D_f = \begin{bmatrix} D \\ D \\ \vdots \\ D \end{bmatrix} \quad (12)$$
$$\overline{G}_f = \begin{bmatrix} 0 & 0 & \cdots & 0 \\ CB_K & 0 & \cdots & 0 \\ \vdots & \vdots & \ddots & \vdots \\ CA_K^{f-2}B_K & CA_K^{f-3}B_K & \cdots & CB_K \end{bmatrix} \quad (13)$$
$$y_f(k) = \begin{bmatrix} y_k \\ y_{k+1} \\ \vdots \\ y_{k+f-1} \end{bmatrix} \quad (14)$$
$$z_{f-1}(k) = \begin{bmatrix} z_k \\ z_{k+1} \\ \vdots \\ z_{k+f-2} \end{bmatrix} \quad (15)$$
By iterating (8) and (9), the following is obtained
$$x_k = \overline{L}_p z_p(k) + A_K^p x_{k-p} \quad (16)$$
where
$$\overline{L}_p = [B_K A_K B_K \cdots A_K^{p-1} B_K] \quad (17)$$
$$z_p(k) = [z_{k-1}^T z_{k-2}^T \cdots z_{k-p}^T]^T \quad (18)$$
By substituting (16) into (11) gives
$$y_f(k) = \Gamma_f \overline{L}_p z_p(k) + \Gamma_f A_K^p x_{k-p} + \overline{G}_f z_{f-1}(k) + D_f u_f(k) + e_f(k) \quad (19)$$
In subspace identification methods, the following assumptions are made: the eigenvalues of $A - KC$ lies inside the unit circle, $(A, C)$ is observable, $(A, [B, K])$ is controllable and $e_k$ is a stationary, zero mean white noise with covariance $E[w_k w_k^T] = R$

Let consider an input vector $x(k)$ and output vector $y(k)$, the linear regression can be expressed as
$$y(k) = \theta x(k) + v(k) \quad (20)$$
which can be written in matrix form as
$$[y(1) y(2) \dots y(N)] = \theta[x(1) \, x(2) \dots x(N)] + V \quad (21)$$
where V is a noise vector.

By minimizing the following function
$$J = \|Y - \theta X\|_F^2 \quad (22)$$
where $\|.\|_F$ is the $F^{th}$ norm, the least square solution is given as
$$\hat{\theta} = YX^T(XX^T)^{-1} \quad (23)$$
and the model prediction is then given as
$$\hat{Y} = \hat{\theta}X = YX^T(XX^T)^{-1}X = Y\Pi x \quad (24)$$
The least square residual is given as
$$\hat{Y} = Y - \hat{Y} = Y(I - \Pi x) = Y\Pi\frac{1}{x} \quad (25)$$
Based form (1) and (2), an extended state space model can be formulated as

$$Y_f = \Gamma_f X_k + H_f U_f + G_f E_f \qquad (26)$$
where $f$ is the future horizon and
$$\Gamma_f = \begin{bmatrix} C \\ CA \\ \vdots \\ CA^{f-1} \end{bmatrix} \qquad (27)$$

$$H_f = \begin{bmatrix} D & 0 & \cdots & 0 \\ CB & D & \cdots & 0 \\ \vdots & \vdots & \ddots & \vdots \\ CA^{f-2}B & CA^{f-3}B & \cdots & D \end{bmatrix} \qquad (28)$$

$$G_f = \begin{bmatrix} I & 0 & \cdots & 0 \\ CK & I & \cdots & 0 \\ \vdots & \vdots & \ddots & \vdots \\ CA^{f-2}K & CA^{f-3}K & \cdots & I \end{bmatrix} \qquad (29)$$

$$U_f = \begin{bmatrix} u_k & u_{k+1} & \cdots & u_{k+N-1} \\ u_{k+1} & u_{k+2} & \cdots & u_{k+N} \\ \vdots & \vdots & \ddots & \vdots \\ u_{k+f-1} & u_{k+f} & \cdots & u_{k+f+N-2} \end{bmatrix} \qquad (30)$$

The Kalman state is estimated from past input and output data based on equation (16) as
$$x_k = \bar{L}_p z_p(k) + A_K^p x_{k-p} \qquad (31)$$
where
$$x_{k-p} == [x_{k-p} \quad x_{k-p+1} \quad \cdots \quad x_{k-p+N-1}] \qquad (32)$$
From equations 32 and 26
$$\begin{aligned} Y_f &= \Gamma_f \bar{L}_p Z_p + H_f U_f + G_f E_f \\ &= H_{fp} Z_p + H_f U_f + G_f E_f \end{aligned} \qquad (33)$$
where $H_{fp} = \Gamma_f \bar{L}_p$

Under open-loop conditions, $E_f$ is uncorrelated to $U_f$ so
$$\frac{1}{N} E_f U_f^T \to 0 \text{ as } N \to \infty \qquad (34)$$
Furthermore, $E_f$ is uncorrelated to $Z_p$ from the Kalman filter theory. Therefore
$$\frac{1}{N} E_f Z_p^T \to 0 \text{ as } N \to \infty \qquad (35)$$
In N4SID we have to eliminate first eliminate $U_f$ by post-multiplying $\Pi \frac{1}{U_f}$ by (33).
$$Y_f \Pi \frac{1}{U_f} = H_{fp} Z_p \Pi \frac{1}{U_f} + H_f U_f \Pi \frac{1}{U_f} + G_f E_f \Pi \frac{1}{U_f} \qquad (36)$$
Then the noise term is removed by multiplying $Z_p^T$ from the result of (35).
$$\begin{aligned} Y_f \Pi \frac{1}{U_f} Z_p^T &= H_{fp} Z_p \Pi \frac{1}{U_f} Z_p^T + G_f E_f Z_p^T \\ &= H_{fp} Z_p \Pi \frac{1}{U_f} Z_p^T \end{aligned} \qquad (37)$$
and
$$\widehat{H}_{fp} = Y_f \Pi \frac{1}{U_f} Z_p^T (Z_p \Pi \frac{1}{U_f} Z_p^T)^{-1} \qquad (38)$$
N4SID performs Singular Value Decomposition (SVD) on
$$\widehat{H}_{fp} Z_p = \Gamma_f L_p Z_p = \widehat{USV^T} \approx U_n S_n V_n^T \qquad (39)$$
where $S_n$ contains the n largest singular value and chooses $\widehat{\Gamma}_f = U_n S_n^{1/2}$.

## IV. MODEL PREDICTIVE CONTROL

Model Predictive Control (MPC) is considered to be an advanced optimized control method that has been widely used in process industries over the last two decades. The strategy that MPC uses to calculate the control actions characterized as: At the $k^{th}$ sampling instant, the values of the control action, $u$, for the next 'M' sampling instants $\{u(k), u(k+1), \ldots, u(k+M-1)\}$ are calculated. They are calculated by minimizing the difference between the predicted outputs and the reference trajectories over the next 'P' sampling instants while satisfying the constraints. In MPC, the control horizon 'M' and the predicted horizon 'P' are the tuning parameters. After calculating the control moves for 'M' sampling instants. The controller will implement the first control move $u(k)$. At the next sampling instant, $k+1$, the control moves are recalculated for the next 'M' sampling instants, [k+1 to k+M], and the first control move $u(k+1)$ is implemented. These steps are repeated at each sampling instants.

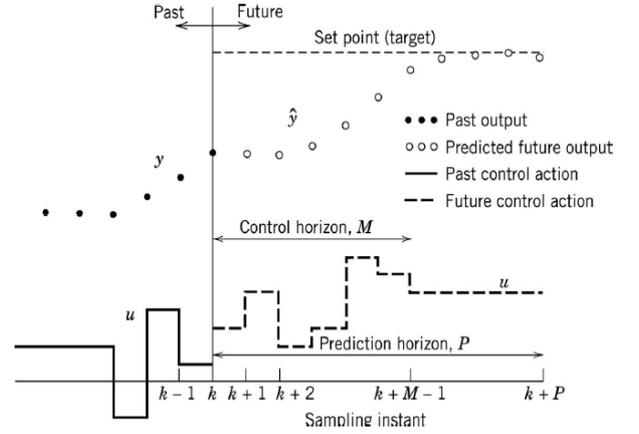

Fig.1 MPC Strategy

In MPC strategy, the dynamical model of the process being controlled is required. It is needed to calculate the future predicted output. The dynamical model of the process must be accurate thus it must capture the dynamics of the processes completely. If the dynamical model of the process is not accurate then the result of the controller might not give the desired performance of the process outputs. There are many different types of models which can be used in MPC, such as impulse response, step response, transfer function models and sate space models.

To obtain the control law, objective function is needed. Different objective function leads to different algorithms. There are several different choices for objective functions. The general one is a standard least-squares or quadratic objective function. The objective function is a sum of squares of the predicted residuals and the control moves. A quadratic objective function can be written as

$$J(k) = \sum_{i=1}^{P} \|\hat{y}(k+i|k) - r(k+i)\|_{Q(i)}^2 \\ + \sum_{i=0}^{M-1} \|\Delta\hat{u}(k+i|k)\|_{R(i)}^2 \quad (40)$$

where $\hat{y}_k$ represents the model predicted output, r is the reference set-point, $\Delta\hat{u}$ is the change in manipulated input from one sample time to the next, R is a weight for the changes in the manipulated input, Q is a weight for the changes in the predicted output, and the subscripts indicate the sample time k. P is a prediction horizon and M is a control horizon.

The controller minimizes the objective function given in equation (40) to obtain the control moves. The objective function is minimized at each sampling instant using quadratic equation solvers.

There are many different quadratic solvers available such as; interior point, active set, augmented Lagrangian, conjugate gradient, KIWIK algorithm, etc. The details of these algorithms are not considered in this paper.

In general the quadratic problem can be formulated as

$$\min_x (f^T u + \frac{1}{2} u^T H u) \quad (41)$$

Subject to $Ax \leq b$

where u is the optimal solution which gives the minimum, H is the positive definite Hessian matrix, A is a matrix of linear constraint coefficients, and b and f are the vectors. The H and A matrices are constants matrices which are calculated during the initialization of the controller and b and f vectors are calculated at the beginning of each sampling instant.

The quadratic solver tries to find the minimum of the function given in (41) which satisfies the constraints.

In MPC calculation, it is assumed that all the states of the process are measurable but that's not true. In most of the application it is impossible to measure all the states and the estimation of the states are required. Observer, Kalman filter and Extended Kalman filter can be used for the state estimation.

## V. PROPOSED ALGORITHM

In this work; the process model is estimated using subspace identification N4SID algorithm. This algorithm gives a good result and it can be extended for MIMO systems. In N4SID algorithm, only the order of the system is the unknown parameter which reduces the complexity of the processes which are estimated. Since the estimated model does not give 100% fit; especially for nonlinear systems having more than one operating conditions, multi-model approach is proposed. In multi model, more than one model in MPC design is used. The MPC will compare the value of objective functions for all the models and select the model which gives the minimum objective function at each sampling instants. After selecting the model of the process, the controller updates the predicted states for every model at that instant. Let consider two models used in MPC design, if the objective function of the first model at sampling instant k is less than the objective function of the second model at that instant. The controller will select the first model and update the predicted model output and states for the second model to be equal to the first model output. It will then implement the control action and repeat the procedure at every sampling instant.

The proposed Algorithm is as follows

**Algorithm Proposed**

**System Identification**
- Apply PRBS signals as input signals to the process based on the prior knowledge of the process. The knowledge could be steady state gain and operating frequencies.
- Obtain input and output data.
- Estimate the models for the process using subspace identification algorithm (N4SID) to estimate the state space matrices.
    - Select the order according to Akaike information criterion (AIC).
    - Estimate two or more models using two different input and output data or by using NH4 option in estimating the state space matrices using N4SID.

**MPC**
- Measure the states x(k|k) and outputs y(k|k) for all models.
    - If states and outputs are not measurable then
        - Estimate the states and outputs using Kalman filter for linear systems
        - Estimate the states and outputs using Extended Kalman filter for nonlinear systems.
- Minimizes the objective functions for all models using quadratic problem solvers.
- Compare the objective functions for all models.
- Select the model which has the minimum objective function.
- Update the predicted states for other models to be equal to the model selected.
- Implement the first control u(k).
- Time update.
- Repeat

## VI. FLUID CATALYTIC CRACKING UNIT

The fluid catalytic cracking unit (FCCU) is complex interactive process in petroleum refining industries. It takes chains of hydrocarbons and breaks them into smaller ones which allow refineries to utilize their crude oil resources more efficiently. It uses an extremely hot catalyst to crack the hydrocarbons into smaller ones. FCCU processes present challenging control problem because it has very complex kinetics of both cracking and coke burning reactions and it has strong interaction between the reactor and regenerator. A schematic of FCCU is shown in fig. 2.

The FCCU contains of two main parts: the reactor and the regenerator. In reactor, the reaction between the mixture of hydrocarbon vapors and catalyst takes place. This reaction breaks the long molecules of hydrocarbons into smaller one which leaves from the top of the reactor. Steam is supplied to remove hydrocarbons from the catalyst. The cracking reaction produces carbon materials and un-cracked organic materials known as coke that reduces the catalyst activity. The catalyst is taken into regenerator where it is regenerated by burning off the deposited coke with air. The regenerated catalyst is then taken to the reactor to repeat the cycle. The combustion of the coke in the regenerator produces a heat absorbed by the regenerated catalyst. This absorbed heat provides the energy for the vaporization and cracking reactions that take place in the catalyst riser.

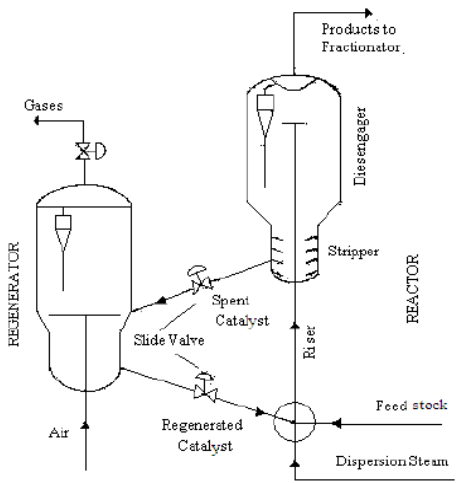

Fig.2 Schematic of FCCU

The FCCU considered here is developed from [Skogested, 1993]. It has two manipulated variable and two controlled variables. The manipulated variables are: $F_s$ which is the flow rate of regenerated catalyst and $F_a$ which is the flow rate of air to regenerator. The controlled variable are: $T_{ro}$ which is the riser outlet temperature and $T_{rg}$ which is the regenerator temperature.

## VII. RESULTS AND DISCUSSIONS

PRBS signals are applied as input signals to the model of the FCCU developed in Simulink to estimate the state space models. This developed from the FCCU model suggested by Skogested in 1993. It has two manipulated variable and two controlled variables. The manipulated variables are: the flow rate of the air to the regenerator $F_a$ and the flow rate of the regenerated catalyst $F_s$. The controlled variable are: riser outlet temperature $T_{ro}$ and the regenerator temperature $T_{rg}$.

5000 samples of input and output data are taken with the sampling rate equals to 0.5 seconds. 2500 samples are used to estimate the models of FCCU and the rest 2500 samples are used for the validation.

The estimated state space model is shown in fig. 3 and fig.4. It produced a fit of 87.83% for the riser outlet temperature, and 84.68% fit for the regenerator temperature.

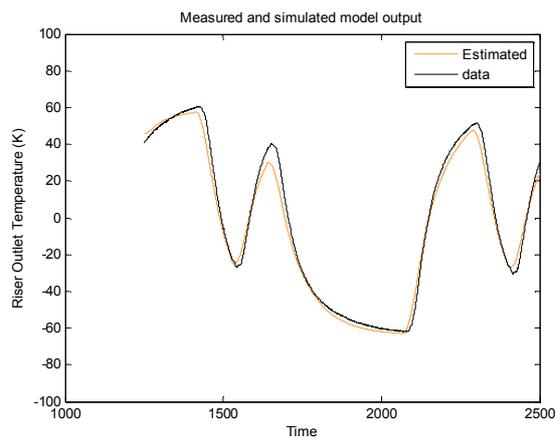

Fig.3 Estimated State Space Model for the Riser Outlet temperature

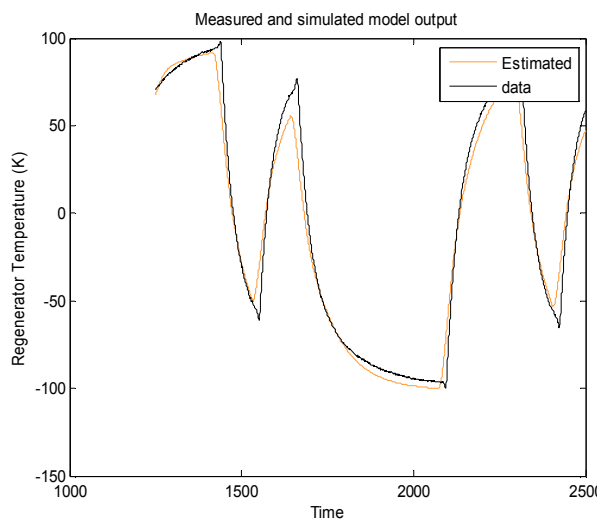

Fig.4 Estimated State Space Model for the Regenerator Temperature

In MPC design, the prediction horizon is chosen to be 50 seconds and the control horizon is chosen to be 25 seconds. The sampling time is chosen to be equal to the sampling time of the estimated models which is 0.5. The input weight is chosen to be equals to zero and the output weight is chosen to be equals to 1. Tracking capability of MPC controlled FCCU system is investigated. The constraints in output variables is chosen to be between [0, 800] for the riser outlet temperature and between [0, 1150] for the regenerator temperature.

In multi model MPC design, two estimated state space models are used. The first model is used N4SID algorithm without any options and the second one used forward and backward prediction. In the second model the maximum forward prediction horizon is chosen to be equals 30, the number of past outputs is chosen to be equals to 45, and the number of past inputs which are used for the predictions is chosen to be equals 45.

The result of the simulation is shown in fig. 5, fig. 6, fig.7 and fig.8. This simulation result shows that the multi model MPC design produces better result than single MPC design. The rising time in multi model MPC design is faster than single

MPC design. The settling time is less in multi model MPC design. The overshoot value is acceptable in the multi model design. This better performance is due to the fact that multi models MPC uses two models and select the one which produce better response at each sampling instant.

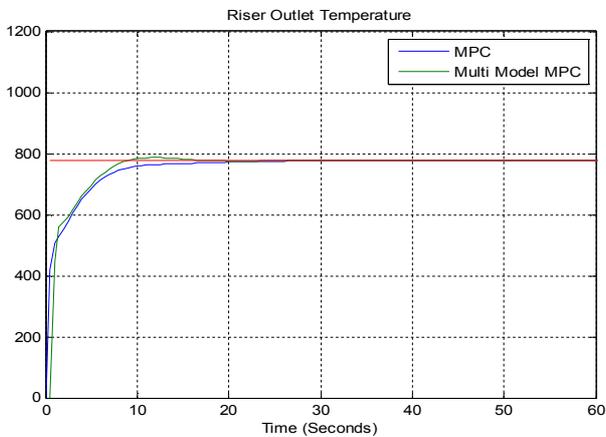

Fig.5 Set-point tracking in single model MPC Vs multi model MPC for the riser outlet temperature for FCCU

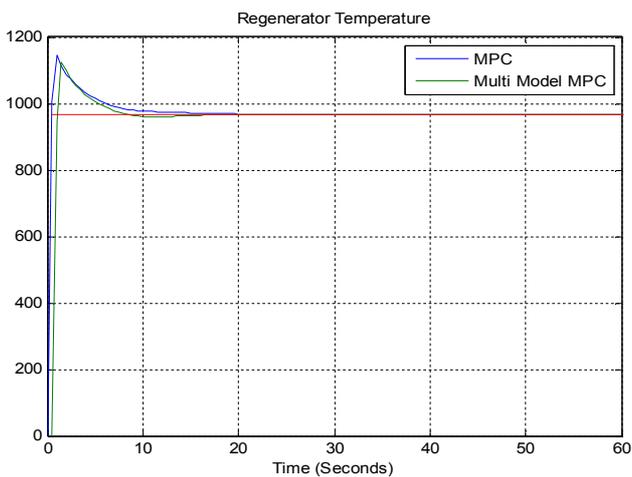

Fig.6 Set-point tracking in single model MPC Vs multi model MPC for the regenerator temperature for FCCU

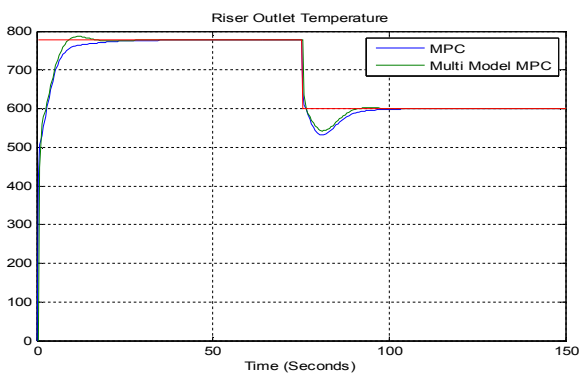

Fig.7 Comparison of set-point tracking in single model MPC Vs multi model MPC for the riser outlet temperature for FCCU

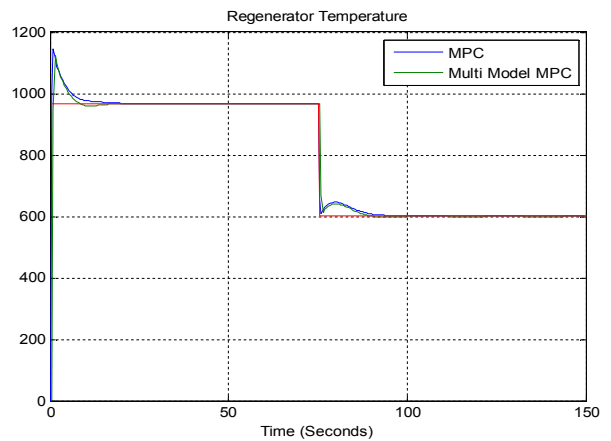

Fig.8 Comparison of set-point tracking in single model MPC Vs multi model MPC for the regenerator temperature for FCCU

Fig. 9, fig. 10, fig. 11 and fig. 12 show the effect of the disturbance on the response of MPC controller design and proposed multi model MPC controller design. The disturbance signal affecting the plant from the time 84 seconds and so on. From these figures, it is clear that multi model MPC gives better disturbance rejection performance.

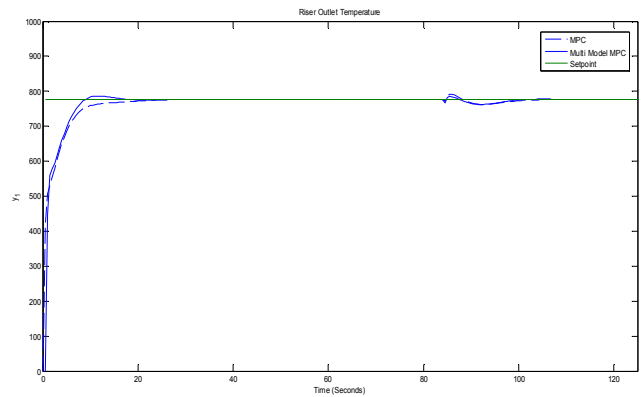

Fig.9 Disturbance rejection for the riser outlet temperature in single model MPC Vs multi model MPC

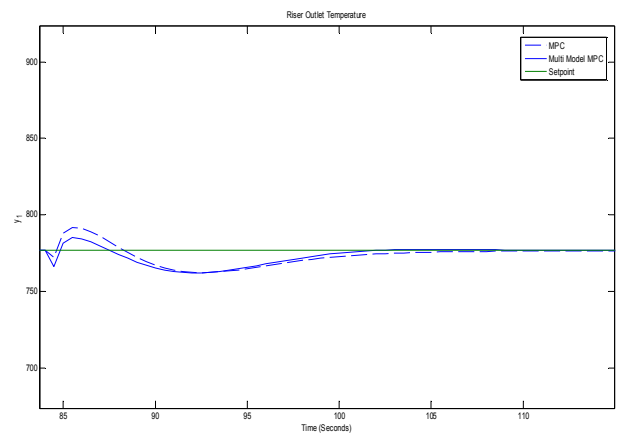

Fig.10 Zoomed view for the disturbance rejection for the riser outlet temperature in single model MPC Vs multi model MPC

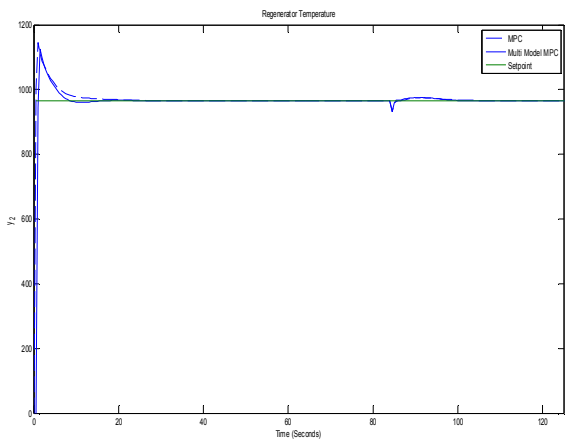

Fig.11 Disturbance rejection for the regenerator temperature in single model MPC Vs multi model MPC

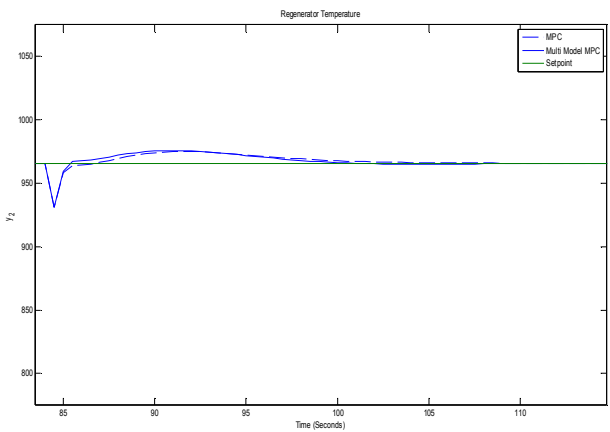

Fig.12 Zoomed view for the disturbance rejection for the regenerator temperature in single model MPC Vs multi model MPC

VIII. CONCLUSION

In this paper, modified multi model predictive control algorithm is proposed. The proposed method is based on system identification techniques and model predictive design. MATLAB/Simulink is used to simulate the result. The simulation results show that the proposed multi model predictive control algorithm provides better result for both reference tracking and disturbance rejection.